\documentclass[a4paper,twocolumn,11pt,unpublished]{quantumarticle}
\pdfoutput=1
\usepackage[utf8]{inputenc}
\usepackage[english]{babel}
\usepackage[T1]{fontenc}
\usepackage{amsmath}
\usepackage{hyperref}
\usepackage{grffile}
\usepackage{threeparttable}
\usepackage{enumitem}
\usepackage{booktabs}%
\usepackage{mathtools}
\usepackage[numbers]{natbib}
\hypersetup{colorlinks,citecolor=blue}

\begin{document}

\title{Emergent Bifurcations in Quantum Circuit Stability from Hidden Parameter Statistics}

\author{Pilsung Kang}
\affiliation{Department of Software Science, Dankook University\\ Yongin 16890, South Korea}
\email{pilsungk@dankook.ac.kr}
\orcid{0000-0001-8882-8686}
\thanks{This work was supported by the National Research Foundation of Korea (NRF) grant funded by the Korea government (MSIT), grant number RS-2023-00244605.}
\maketitle

\begin{abstract}
The compression of quantum circuits is a foundational challenge for near-term quantum computing, yet the principles governing circuit stability remain poorly understood. We investigate this problem through a large-scale numerical analysis of 300 structurally-uniform circuits across 10, 12, and 14 qubits. Despite their macroscopic uniformity, we find that each ensemble universally bifurcates into distinct robust and fragile classes. We solve the puzzle of this emergent bifurcation, demonstrating that its origin is not structural, but is instead encoded in the statistical properties of the gate rotation parameters. Fragile circuits consistently exhibit a universal signature of ``statistical brittleness,'' characterized by low parameter variability and a scarcity of small-angle gates. We uncover the underlying physical mechanism for this phenomenon: Paradoxical Importance where smaller-angle gates are counter-intuitively more critical to the circuit's function, an effect most pronounced in fragile circuits. This reliance on fine-tuning explains why statistically brittle circuits are uniquely vulnerable to failure under compression. These findings establish a new framework for engineering resilient quantum algorithms, shifting the focus from macroscopic structure to the microscopic statistical properties of a circuit's parameters.
\end{abstract}

\section{Introduction}\label{s:intro}

Optimizing quantum circuits by reducing their gate count and depth is a central challenge in the Noisy Intermediate-Scale Quantum (NISQ) era~\cite{preskill:2018:nisq,bharti:2022:nisq}. As algorithms grow in complexity, such compression techniques become essential to mitigate decoherence and gate errors~\cite{kokcu:2022:aps,gibbs:2025:deepcircuit}. The prevailing assumption has been that robustness under compression is primarily determined by macroscopic structural properties like gate count and depth. Consequently, circuits sharing identical macroscopic features are expected to exhibit similar resilience.

However, this assumption masks a deeper and more complex reality. We performed a large-scale numerical study on 300 structurally-uniform circuits across 10, 12, and 14 qubits and found this assumption to be false. We discovered a universal phenomenon: at every scale, the ensembles bifurcated into distinct ``robust'' and ``fragile'' classes despite their macroscopic uniformity. This represents a critical knowledge gap: if not the high-level structure, what is the fundamental mechanism that governs a circuit's stability?

In this work, we solve this puzzle by proposing the \textbf{Hidden Landmine Hypothesis}, where fragility is not caused by specific structural defects but emerges from a universal statistical property of the circuit's rotation parameters, a state we term ``statistical brittleness.'' This work establishes the microscopic statistical properties of a circuit's parameters as the definitive factor in its stability, providing a new, scale-aware framework for designing and optimizing resilient quantum algorithms.

\section{Methodology}\label{s:setup}

\begin{table*}[htbp!]
\centering
\caption{Key parameters for the three uniform circuit ensembles. The systematic increase in depth factor and decrease in redundancy rate were designed to maintain a challenging compression regime as the system size scaled.}
\label{tab:exp_params}
\begin{tabular*}{0.9\textwidth}{@{\extracolsep{\fill}}lccc}
\toprule
\textbf{Parameter} & \textbf{10 Qubits} & \textbf{12 Qubits} & \textbf{14 Qubits} \\
\midrule
\textit{Circuit Generation} & & & \\
Number of circuits & 100 & 100 & 100 \\
Number of gates & 342 & 537 & 877 \\
Depth & 46 & 61 & 84 \\
Depth factor ($\alpha$) & 2.3 & 2.5 & 3.0 \\
Redundancy rate ($\rho$) & 0.28 & 0.25 & 0.2 \\
\midrule
\textit{Analysis Protocol} & & & \\
Compression ratio ($\kappa$) & 0.11 & 0.10 & 0.08 \\
Fidelity metric & \multicolumn{3}{c}{$F = |\langle\psi_{\text{orig}}|\psi_{\text{compressed}}\rangle|^2$} \\
Importance metric & \multicolumn{3}{c}{$I_i = 1 - F(\mathcal{C}_{\setminus i})$} \\
\bottomrule
\end{tabular*}
\end{table*}

Our study is based on a comprehensive numerical analysis designed to isolate the structural origins of circuit robustness. The key parameters defining the scope and methodology of our experiments are summarized in Table~\ref{tab:exp_params}.  All numerical experiments, including circuit generation, state vector simulation, and fidelity analysis, were implemented using the Qiskit open-source framework~\cite{javadiabhari:2024:qiskit}. The source code and data that support the findings of this study are openly available in a GitHub repository at \url{https://github.com/pilsungk/Q-Compress}.

\subsection{Quantum Circuit Construction}

To isolate the effects of microscopic arrangements, we generated ensembles of structurally-uniform parametrized quantum circuits. Each circuit with $n$ qubits is constructed from a number of layers given by $\lfloor n \cdot \alpha \rfloor$, where $\alpha$ is a tunable depth factor. A single layer consists of a sub-layer of single-qubit rotations ($R_{\text{axis}}$ where $\text{axis} \in \{X, Y, Z\}$) applied to all qubits, followed by an entangling sub-layer of CNOT gates applied to alternating pairs of adjacent qubits to form a brick-wall structure ensuring full nearest-neighbor connectivity, with this entangling sub-layer being omitted from the circuit's final layer.

The crucial parameter, the \textbf{redundancy rate} ($\rho$), serves two functions in our circuit generation. First, it determines the probability that a rotation gate within the layered ansatz is assigned a near-zero angle ($\theta \in [0.001, 0.05]$) versus a significant angle ($\theta \in [\pi/6, \pi/2]$). Second, after the main layers are constructed, a number of additional single-qubit $R_z(\theta)$ gates, equal to $\lfloor n \cdot \rho \rfloor$, are appended to randomly selected qubits. These appended gates are also assigned a near-zero angle ($\theta \in [0.001, 0.01]$), further contributing to the circuit's overall redundancy.  To generate statistically independent instances for our uniform ensembles, we used a deterministic pseudo-random number generator seeded differently for each circuit. This ensures that all circuits in an ensemble share the same macroscopic properties (total gate count, depth, and target redundancy rate $\rho$) while varying only in their microscopic details (the specific assignment of rotation axes and angles). This methodology allows us to attribute any observed differences in robustness directly to the microscopic gate arrangement, rather than to architectural confounding factors.

\subsection{Compression via Causal Importance Ranking}

Our compression protocol, which we term \textbf{Causal Pruning}, was designed to remove gates based on their functional importance. The causal importance $I_i$ of each gate $i$ was quantified using a leave-one-out analysis: $I_i = 1 - F(\mathcal{C}_{\setminus i})$, where $F$ is the fidelity and $\mathcal{C}_{\setminus i}$ is the circuit with gate $i$ removed. Once the importance of all gates was computed, they were ranked in ascending order, and the least important gates were sequentially pruned to achieve a target compression ratio $\kappa$.

\subsection{Importance Distribution Analysis}
All fidelity calculations were performed via exact state vector simulation. After calculating the full set of importance scores for each circuit, we characterized the shape of the resulting distribution to test various structural hypotheses. Our primary metric for this was the Shannon entropy of the normalized importance scores, $H = -\sum_i p_i \ln(p_i)$. Here, $p_i$ is the normalized importance of the $i$-th gate, defined as its individual importance $I_i$ divided by the sum of all importance scores. This metric quantifies the uniformity of the distribution, where a low entropy indicates a hierarchical structure, and a high entropy suggests a more uniform distribution of importance across the gates.

\subsection{Identifying the Critical Transition Regime}

To systematically identify the compression ratio $\kappa$ that exhibits the clearest robust/fragile transition for each qubit system, we employed a targeted parameter search strategy.

\textbf{Circuit Parameter Selection:} The circuit generation parameters ($\alpha$ and $\rho$) were determined using empirically-derived heuristics based on the target qubit count. For our experiments, we used $\alpha$ values ranging from 2.3 to 3.0 and $\rho$ values from 0.20 to 0.28, with larger systems generally requiring higher $\alpha$ and lower $\rho$ to maintain circuit complexity while ensuring sufficient gate diversity. While these parameters were held fixed for each qubit system in our current study, the framework supports systematic sweeps over these parameters when needed for broader parameter space exploration.

\textbf{Compression Ratio Search:} Using the empirically-determined circuit parameters ($\alpha, \rho$), we performed a systematic sweep of $\kappa$ from 0.05 to 0.40 in 0.03 increments. For each value of $\kappa$, we generated an ensemble of 30 probe circuits and evaluated the separation quality between robust (fidelity $\geq$ 0.9) and fragile (fidelity $<$ 0.9) populations. The optimal $\kappa$ was selected based on maximizing the fidelity gap between these two classes, ensuring the clearest demonstration of the robustness transition phenomenon.

This process allowed us to identify the specific, critical parameter regime (detailed in Table~\ref{tab:exp_params}) where the underlying ``Hidden Landmine'' physics is most clearly manifested.

\section{Results}\label{s:results}

In this section, we present the results of our investigation in three stages. We first demonstrate the central macroscopic phenomenon: the emergent bifurcation of structurally-uniform circuits into robust and fragile classes. We then identify the microscopic origin of this bifurcation---a universal statistical signature in the circuits' rotation parameters. Finally, we uncover the underlying physical mechanism of this signature, a paradoxical fine-tuning, that provides a complete explanation for the observed fragility.

\subsection{The Macroscopic Puzzle: Identical Structures, Divergent Fates}

\begin{figure}[htbp!]
\centering
\includegraphics[width=0.95\columnwidth]{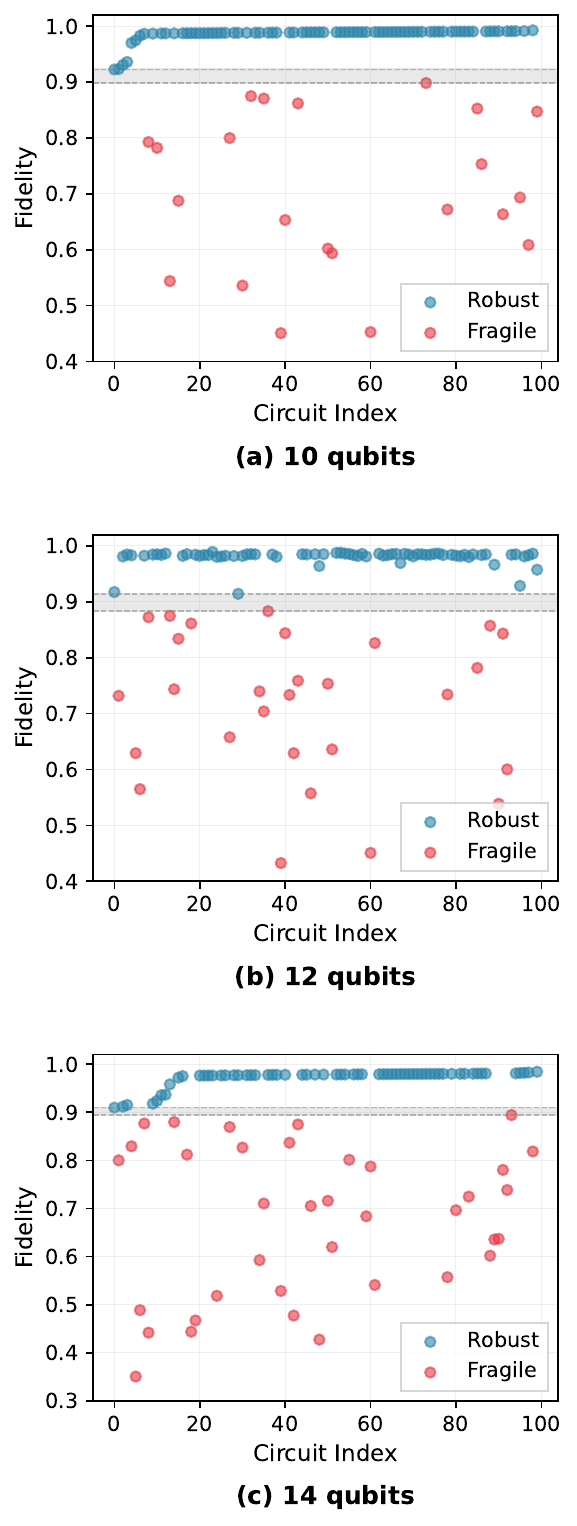}
\caption{Universal bimodal separation across quantum circuit scales. (a-c) Fidelity distributions for 10, 12, and 14 qubit circuits under compression ratios ($\kappa$) of 11\%, 10\%, and 8\% respectively. Despite identical macroscopic parameters, circuits bifurcate into robust (blue) and fragile (red) classes. The large effect size, quantified by Cohen's d, confirms the statistical significance of the separation at each scale.}
\label{fig:fidelity_separation}
\end{figure}

Our investigation reveals a striking and universal phenomenon across quantum circuits of varying scales. We generated three ensembles of 100 structurally-uniform circuits each at 10, 12 and 14 qubits, with every circuit within an ensemble sharing identical macroscopic parameters---fixed gate counts, depths, and connectivity patterns. Conventional wisdom suggests that such structurally identical circuits should exhibit similar robustness under compression.

Contrary to this expectation, all three ensembles exhibit perfect bimodal separation when subjected to compression. As illustrated in Figure~\ref{fig:fidelity_separation} and detailed in Table~\ref{tab:experiment_overview}, circuits cleanly bifurcate into two distinct classes regardless of system size: a robust class maintaining high fidelity and a fragile class suffering significant performance degradation. This separation is absolute, with clear fidelity gaps of 0.0242, 0.0311, and 0.0154 for 10, 12, and 14 qubits respectively. The magnitude of this separation is confirmed by the exceptionally large effect sizes (Cohen's d $\approx$ 4.39, 3.88, and 3.15), indicating a statistically unambiguous bifurcation.

The universality of this phenomenon across scales presents a profound puzzle. Within each ensemble, circuits are identical by all conventional metrics---gate counts, depths, and entanglement ratios remain constant. Yet their responses to compression diverge dramatically. This finding directly challenges the paradigm that macroscopic parameters are sufficient to predict circuit stability. 

\begin{table}[htbp!]
\centering
\caption{Experimental overview and bimodal separation across scales. For each system size, 100 structurally-uniform circuits were analyzed, showing the compression ratio ($\kappa$) applied, the resulting classification, mean fidelities, and the fidelity gap.}
\label{tab:experiment_overview}
\begin{tabular*}{\columnwidth}{@{\extracolsep{\fill}}l ccc}
\toprule
\textbf{Metric} & \textbf{10q} & \textbf{12q} & \textbf{14q} \\
\midrule
Comp. Ratio ($\kappa$) & 11\% & 10\% & 8\% \\
Robust Circuits & 78\% & 72\% & 63\% \\
Fragile Circuits & 22\% & 28\% & 37\% \\
\midrule
Robust Fidelity & 0.985 & 0.980 & 0.972 \\
Fragile Fidelity & 0.704 & 0.717 & 0.675 \\
Fidelity Gap & 0.0242 & 0.0311 & 0.0154 \\
\bottomrule
\end{tabular*}
\end{table}

\subsection{The Microscopic Culprit: A Statistical Signature of Fragility}

The stark bifurcation in circuit fidelity begs the question: what microscopic property distinguishes the fragile circuits from the robust ones? While initial analyses of high-level statistical metrics of gate importance, such as entropy or the Gini coefficient, revealed no statistically significant differences, a deeper investigation into the fundamental gate parameters ($\vec{\theta}$) uncovered the true origin of fragility. Our analysis revealed a consistent and statistically significant fingerprint that reliably identifies fragile circuits across all tested system sizes.

\begin{table*}[htbp!]
\centering
\caption{Key statistical differences in gate parameter distributions between robust and fragile classes across all tested scales. The statistical fingerprint of fragility—higher mean, lower standard deviation, and fewer small-angle gates—is consistently observed, confirming the universality of the mechanism.}
\label{tab:param_stats_combined}
\begin{tabular*}{\textwidth}{@{\extracolsep{\fill}}lcccccc}
\toprule
\textbf{Rotation Parameter ($\theta$)} & \multicolumn{2}{c}{\textbf{10 Qubits}} & \multicolumn{2}{c}{\textbf{12 Qubits}} & \multicolumn{2}{c}{\textbf{14 Qubits}} \\
\cmidrule(lr){2-3} \cmidrule(lr){4-5} \cmidrule(lr){6-7}
                          & Robust & Fragile & Robust & Fragile & Robust & Fragile \\
\midrule
Mean Angle ($\bar{\theta}$)      & 0.743  & 0.778   & 0.780  & 0.803   & 0.834  & 0.855   \\
Angle Std. Dev. ($\sigma_{\theta}$) & 0.531  & 0.520   & 0.519  & 0.511   & 0.495  & 0.483   \\
Small Angle Ratio         & 0.297  & 0.263   & 0.263  & 0.241   & 0.209  & 0.188   \\
\bottomrule
\end{tabular*}
\end{table*}

This statistical brittleness is characterized by a distinct set of properties in the distribution of rotation angles. As detailed in Table~\ref{tab:param_stats_combined}, fragile circuits consistently exhibit:

\begin{itemize}[itemsep=1pt]
    \item A higher mean rotation angle ($\bar{\theta}$).
    \item A lower standard deviation of rotation angles ($\sigma_{\theta}$), indicating less diversity.
    \item A smaller proportion of near-zero angle gates.
\end{itemize}

The fact that this statistical signature persists and remains highly significant across 10, 12, and 14-qubit systems provides definitive evidence that it is the fundamental mechanism driving the observed stability bifurcation.

\begin{table}[htbp!]
\centering
\caption{Scale-dependence of contributing gate types. The table shows the p-values for the statistical comparison of rotation parameter ($\theta$) distributions between robust and fragile classes, broken down by gate type. The signature of fragility is primarily driven by $R_y$ and $R_z$ gates at 10 qubits, but shifts to $R_x$ and $R_y$ gates at larger scales. Statistically significant results ($p < 0.05$) are highlighted in bold.}
\label{tab:gate_type_dependence}
\begin{tabular*}{\columnwidth}{@{\extracolsep{\fill}}cccc}
\toprule
\textbf{Gate Type} & \textbf{10q} & \textbf{12q} & \textbf{14q} \\
\midrule
$R_x(\theta)$ & 0.095 & \textbf{0.004} & \textbf{<0.001} \\
$R_y(\theta)$ & \textbf{0.013} & \textbf{0.002} & \textbf{0.002} \\
$R_z(\theta)$ & \textbf{0.003} & 0.446 & 0.115 \\
\bottomrule
\end{tabular*}
\end{table}

A deeper look into the data, summarized in Table~\ref{tab:gate_type_dependence}, reveals a further subtlety. The statistical signature found in the rotation parameter ($\theta$) distributions of fragile circuits is universal. However, the specific gate types ($R_x$, $R_y$, $R_z$) responsible for this signature change with the system size. In the 10-qubit system, the statistically significant difference is most pronounced in the $R_y(\theta)$ and $R_z(\theta)$ gates. As the system scales to 12 and 14 qubits, however, the signature shifts to being primarily driven by the $R_x(\theta)$ and $R_y(\theta)$ gates, while the $R_z(\theta)$ gates no longer show a statistically significant difference. This suggests a complex interplay between the universal fragility mechanism and the system's specific scale and topology.

This leads to a refined understanding of the Hidden Landmine Hypothesis. The landmines are not specific gates, but rather a collective statistical property of the circuit's parameters. A state of low parameter variability and a scarcity of small-angle gates creates a brittle parameter landscape. In this regime, the circuit is less adaptable to perturbations like gate removal, leading to the observed fragility. The mechanism was hidden not because it was structural, but because it required a precise statistical lens to distinguish the subtle, yet critical, differences in the rotation parameter ensembles.

\subsection{The Paradox of Fine-Tuning: An Emergent Signature of Fragility}

To uncover the final link between the statistical properties of parameters and circuit failure, we investigated the relationship between the angle of a rotation gate and its causal importance. The analysis revealed a counter-intuitive physical mechanism that serves as the ultimate explanation for fragility: a phenomenon we term \textbf{Paradoxical Importance}.

Across all system sizes, we observed a negative correlation between gate angle and importance, meaning smaller angles are, paradoxically, more critical to the circuit's function. The statistical significance of this signature, as summarized in Table~\ref{tab:correlation_summary}, shows a remarkable scale-dependence. The clearest evidence emerged in the 12-qubit ensemble (p=0.002), suggesting this scale represents a ``sweet spot'' where the reliance on fine-tuning is most easily detectable. At other scales (10q and 14q), the trend persists but its statistical clarity is submerged, suggesting more complex interactions may obscure this simple correlation.

\begin{table}[htbp!]
\centering
\caption{Angle-Importance correlation across system sizes. The table shows the mean Pearson correlation coefficient (r) between rotation angle and gate importance for the robust and fragile ensembles. A more negative r indicates a stronger paradoxical importance of small-angle gates. This effect becomes statistically significant in the 12-qubit ensemble, highlighting an emergent, scale-dependent signature of fragility.}
\label{tab:correlation_summary}
\begin{tabular*}{\columnwidth}{@{\extracolsep{\fill}}lccc}
\toprule
\textbf{Qubits} & \textbf{\begin{tabular}[c]{@{}c@{}}Robust\\Mean r\end{tabular}} & \textbf{\begin{tabular}[c]{@{}c@{}}Fragile\\Mean r\end{tabular}} & \textbf{p-value} \\
\midrule
10q & -0.271 & -0.302 & 0.097 \\
12q & -0.272 & -0.314 & \textbf{0.002} \\
14q & -0.266 & -0.284 & 0.122 \\
\bottomrule
\end{tabular*}
\end{table}

\begin{figure}[htbp!]
    \centering
    \includegraphics[width=0.97\columnwidth]{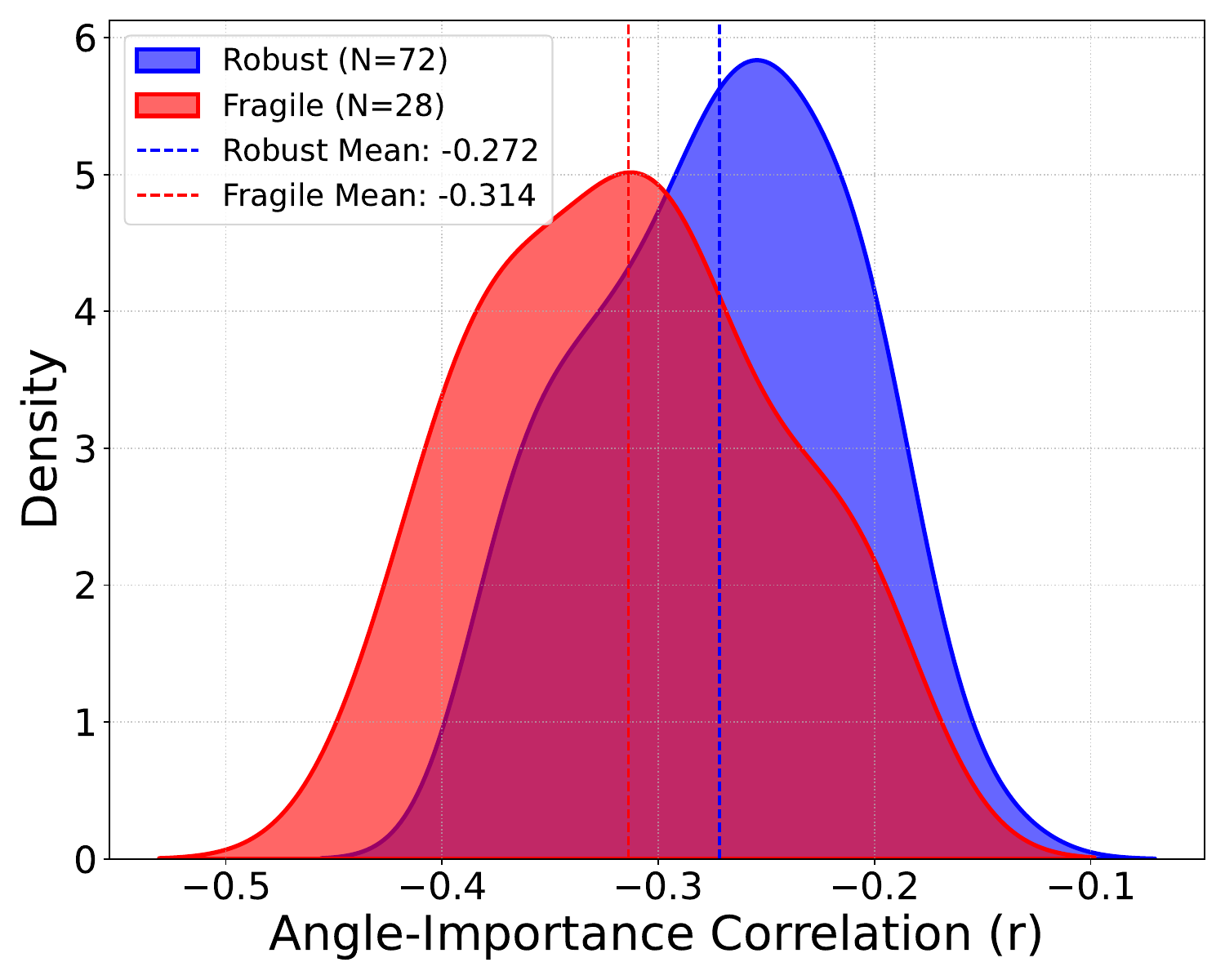}
    \caption{Distribution of the angle-importance correlation coefficient (r) for the full 12-qubit ensemble. The distribution for the fragile circuits (red, N=28) is systematically shifted toward more negative values compared to the robust circuits (blue, N=72), confirming that a stronger paradoxical importance is a defining characteristic of the fragile class.}
    \label{fig:correlation_distribution}
\end{figure}

This statistical tendency at the 12-qubit scale is not a simple shift in mean but a feature of the entire ensemble. Figure~\ref{fig:correlation_distribution} shows the complete probability distributions of the correlation coefficient (r) for the robust and fragile classes. The distribution for fragile circuits is clearly shifted toward a more negative correlation, providing definitive evidence that a stronger Paradoxical Importance is the key statistical signature of fragility.

\begin{figure}[htbp!]
\centering
\includegraphics[width=0.98\columnwidth]{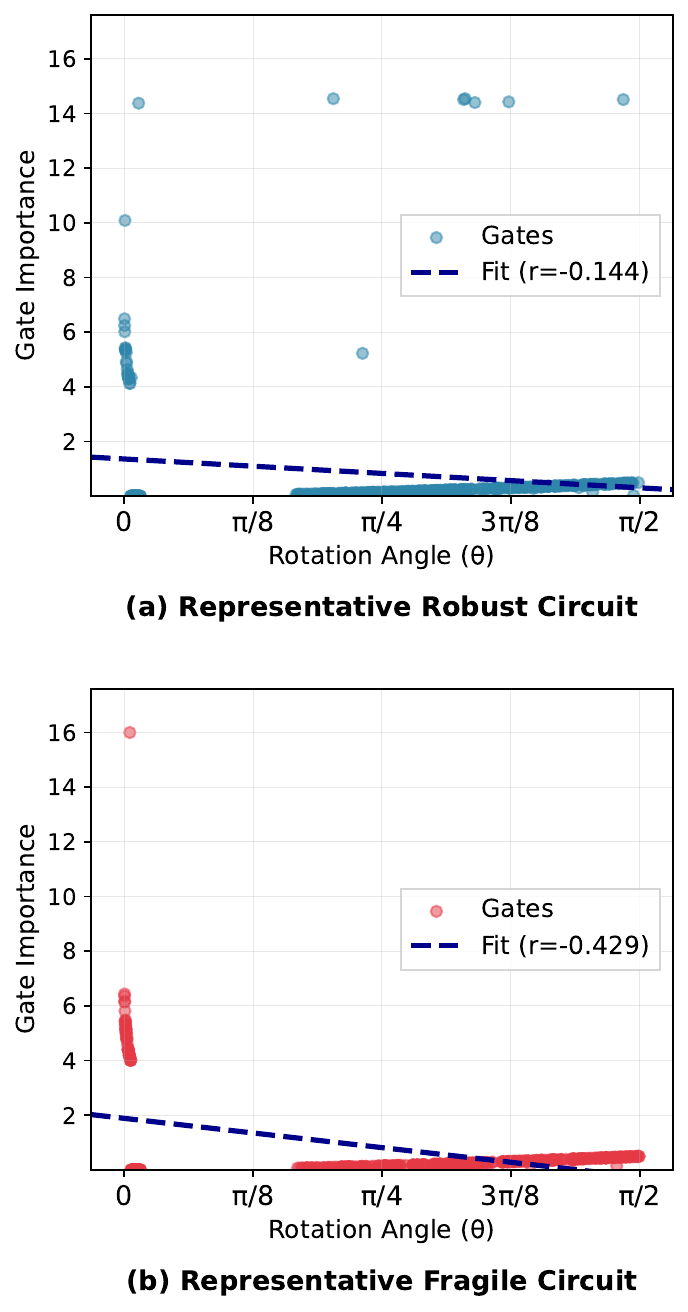}
\caption{Visual evidence of Paradoxical Importance in representative 12-qubit circuits. Each point represents a single rotation gate. (a) In a robust circuit with one of the weakest correlations in its class, the relationship between angle and importance is very weak (r = -0.144). (b) In a fragile circuit with one of the strongest correlations, the paradoxical relationship is visually striking (r = -0.429), demonstrating its critical reliance on the fine-tuning of small-angle gates.}
\label{fig:paradoxical_importance}
\end{figure}

This effect is also visually apparent in individual circuits. Figure~\ref{fig:paradoxical_importance} compares representative robust and fragile circuits from the ensemble. While both exhibit a negative correlation, the trend is significantly stronger and more pronounced in the fragile circuit, providing an intuitive illustration of its reliance on fine-tuning.

The physical mechanism of Paradoxical Importance provides the ultimate explanation for the observed bifurcation. Statistically brittle circuits are those that rely heavily on fine-tuning by these small-angle gates, leaving them uniquely vulnerable to collapse when a crucial gate is removed by the compression algorithm.
\section{Discussion}\label{s:discuss}

Our findings challenge the conventional wisdom that quantum circuit stability is determined by macroscopic features like gate count or depth. We have demonstrated that the origin of fragility lies at a much deeper level: the statistical distribution of the circuit's fundamental rotation-angle parameters ($\vec{\theta}$). This work establishes a new framework where a circuit's vulnerability is not a random occurrence but a predictable consequence of its parameters occupying a statistically brittle region of configuration space. This perspective shifts the focus of circuit design and optimization from simple gate counting to managing the statistical properties of a circuit's core components.

\subsection{The Hidden Landmine Hypothesis Refined}

Our findings culminate in a refined understanding of our central thesis, the Hidden Landmine Hypothesis. The landmine that causes catastrophic failure is not a structural defect, but a subtle, collective property of the circuit's rotation parameters. We have shown that fragility is a direct consequence of statistical brittleness—a state characterized by low parameter variability and a scarcity of small-angle gates. 

The physical mechanism is Paradoxical Importance, where the fine-tuning of these small-angle gates becomes counter-intuitively critical to the circuit's function. This reliance creates an ``information bottleneck'', a concept first introduced in information theory~\cite{tishby:2000:ib,tishby:2015:dl_ib}, where the circuit's logic becomes critically dependent on a few key gates, making the entire system vulnerable. When a compression algorithm, even a perfect one like Causal Pruning, is forced to remove one of these seemingly innocuous gates, it severs this bottleneck and triggers a cascade failure. The mechanism was hidden not because it was structurally invisible, but because its origin lies in the subtle, global statistics of the parameter ensemble, a property to which conventional circuit analysis is blind. This provides a complete physical picture, from microscopic parameter statistics to macroscopic circuit failure.

\subsection{Theoretical Implications: Statistical Brittleness and a Phase-Transition-Like Phenomenon}

The observed bifurcation into robust and fragile classes is strongly reminiscent of a first-order phase transition in statistical mechanics, drawing parallels to the robust-yet-fragile nature of other complex networks~\cite{carlson:2002:pnas}. The key components of such a transition can be mapped directly to our system:

\begin{itemize}[itemsep=3pt]
    \item \textbf{Two Phases:} The system exhibits two distinct phases: a robust phase characterized by high fidelity under compression, and a fragile phase characterized by a catastrophic collapse in fidelity.
    
    \item \textbf{Control Parameter:} The compression ratio ($\kappa$) acts as an external control parameter that drives the system between regimes. Our observation of a critical $\kappa$ for the onset of fragility further strengthens this analogy.

    \item \textbf{Order Parameter:} The statistical brittleness of the circuit, quantified by the statistical properties of the rotation parameters (e.g., low $\sigma_{\theta}$, high $\bar{\theta}$), serves as the hidden order parameter. This intrinsic property determines which phase a circuit belongs to, even before compression is applied.
    
    \item \textbf{Discontinuous Jump:} For a fragile circuit, increasing the control parameter $\kappa$ beyond a critical threshold results in a discontinuous jump in the system's state (i.e., a sudden drop in fidelity), which is a hallmark of a first-order transition.
\end{itemize}

While not a true thermodynamic transition, this framework is powerful. It suggests that circuit stability can be understood through the lens of critical phenomena, where the microscopic statistical state of the parameters dictates the macroscopic, emergent behavior of the system under external stress ($\kappa$).

\subsection{Practical Applications and Design Principles}
Our findings translate directly into a new set of data-driven principles for designing and optimizing robust quantum circuits. By shifting the focus from structural properties to the statistical ensemble of gate parameters, we can move from post-mortem analysis to proactive, robustness-aware design.

A core design principle is to actively manage the statistical landscape of the circuit's parameters. The strongest indicator of fragility is low parameter variability ($\sigma_{\theta}$), so techniques that encourage a wider spread of rotation angles could significantly enhance resilience. Similarly, our work suggests that small-angle gates, often aggressively removed during optimization, may act as a softening agent, providing the parametric flexibility needed to absorb perturbations. Their strategic preservation is therefore crucial for robustness.

These principles also pave the way for a new generation of \textbf{Statistically-Aware Compression} algorithms. We propose a two-phase framework where the compiler first performs a rapid statistical risk assessment of the circuit's parameters (e.g., $\bar{\theta}$, $\sigma_{\theta}$). Based on this assessment, the compression strategy is then adapted. For circuits flagged as brittle, the algorithm would employ a more cautious approach, such as limiting the removal of small-angle gates or prioritizing optimizations that increase parameter diversity. This represents a paradigm shift from simple gate-count reduction to a more holistic, reliability-focused optimization strategy, with the potential to significantly improve the performance of algorithms on real-world NISQ devices.

\subsection{Broader Impact and Future Directions}

Our discovery that circuit stability is dictated by the statistical properties of its parameters opens several new avenues for research and has broad implications for the field. For instance, the concept of statistical brittleness extends naturally to the domain of variational quantum algorithms (VQAs)~\cite{peruzzo:2014:vqe}. The classical optimization loop in a VQA, guided only by the expectation value of the Hamiltonian, could inadvertently steer the circuit's parameters into a fragile regime. Our findings suggest a path to mitigate this by introducing a \textbf{Robustness-Aware Regularization} term to the VQA cost function.

For example, a modified cost function $C'(\vec{\theta}) = \langle H \rangle_{\vec{\theta}} + \lambda \cdot f(\sigma_{\theta})$ could be used, where $\langle H \rangle$ is the original objective, $\sigma_{\theta}$ is the standard deviation of the parameters, and $f$ is a penalty function that grows as $\sigma_{\theta}$ decreases. This regularization would explicitly penalize the optimizer for reducing parameter diversity, actively preventing the VQA from entering a statistically brittle state. This suggests a new direction in VQA design, where the resilience of the ansatz is co-optimized alongside the primary objective.

Similarly, in quantum error correction (QEC), our findings suggest a new factor to consider in fault-tolerant design, moving beyond standard error models. QEC performance thresholds are often calculated assuming a simple, uncorrelated physical error model. However, our work indicates that a circuit in a statistically brittle state is prone to cascade failures, a form of highly correlated error. The statistical properties of the physical gates implementing a logical qubit may therefore influence the logical error rate in non-trivial ways.

This opens several new possibilities. For instance, the statistical properties of a physical circuit block could be used as an input to a more sophisticated QEC decoder. A decoder aware of statistically brittle zones could assign a higher probability to correlated error events in those regions, potentially improving decoding accuracy. Furthermore, avoiding statistical brittleness could become a new design principle for fault-tolerant logical gates, where the physical gate sequence is co-optimized to be both correct and robust against this newly identified failure mode.

This work also presents clear directions for future investigation. A key next step is to perform systematic compression ratio sweeps to map out the complete stability phase diagram $F(n, \kappa)$, where $n$ is the number of qubits and $\kappa$ is the compression ratio, which would clarify the nature of the observed bifurcation.  Furthermore, understanding why the specific gate types contributing to the statistical signature change with system size ($R_x(\theta)$/$R_y(\theta)$ for 12 qubits vs. $R_y(\theta)$/$R_z(\theta)$ for 10 qubits) could reveal deeper insights into the interplay between gate function and circuit topology. Ultimately, our work suggests that statistical brittleness could be incorporated into circuit complexity theory as a new, practical measure of a circuit's resilience, providing a more holistic understanding of the resources required for reliable quantum computation.

\section{Conclusion}\label{s:conc}

In this work, we have fundamentally challenged the conventional understanding of quantum circuit stability. We demonstrated that for structurally-identical circuits, resilience under compression is not governed by macroscopic metrics but by the microscopic statistical properties of the gate rotation parameters. We have shown that circuits universally bifurcate into robust and fragile classes, and we have uncovered the definitive signature of this fragility: a state of statistical brittleness characterized by low parameter variability and a scarcity of small-angle gates.

The physical origin of this phenomenon was traced to a counter-intuitive Paradoxical Importance, where smaller-angle gates are disproportionately critical to the circuit's function. This reliance on fine-tuning makes statistically brittle circuits uniquely vulnerable to failure, as even an ideal compression algorithm is forced to remove gates that are essential for maintaining the circuit's delicate balance. This work provides a complete mechanistic picture, connecting the microscopic parameter statistics to the macroscopic, emergent stability of the entire system.

Our findings offer a new perspective on the design and optimization of quantum algorithms. This suggests the focus can be expanded from simply minimizing gate counts to actively managing the statistical landscape of a circuit's parameters. This opens several compelling avenues for future research, from developing robustness-aware compilers that can detect and mitigate statistical brittleness, to mapping the complete stability phase diagram of these systems. By establishing that resilience is an emergent and predictable property of a circuit's microscopic parameters, this work lays a new foundation for engineering robust, large-scale quantum algorithms.


\end{document}